\documentclass[draft,jgrga]{agutexSI}

\usepackage{amssymb}
\usepackage{array,multirow}
 \usepackage{float}
 \usepackage{graphicx}
\usepackage{amsthm}
\usepackage{amsmath}
\usepackage{longtable}
\usepackage{lipsum}
\usepackage{gensymb}
\usepackage{blindtext}
\usepackage{caption}
\usepackage{grffile}
\usepackage{xcolor,soul}

\setkeys{Gin}{draft=false}
\authorrunninghead{YU ET AL.}

\titlerunninghead{Interparticle Forces Measurements of `Tholin' Using AFM}

\authoraddr{Corresponding author: Xinting Yu,
Department of Earth and Planetary Sciences, Johns Hopkins University,
3400 North Charles Street, Baltimore, MD 21218, USA.
(xyu33@jhu.edu)}

\begin{document}

\title{Direct Measurement of Interparticle Forces of Titan Aerosol Analogs (`Tholin') Using Atomic Force Microscopy}

\authors{Xinting Yu\altaffilmark{1}, Sarah M. H\"orst\altaffilmark{1}, Chao He\altaffilmark{1}, Patricia McGuiggan\altaffilmark{2}, Nathan T. Bridges\altaffilmark{3,$\dagger$}}

\altaffiltext{1}{Department of Earth and Planetary Sciences, Johns Hopkins University, Baltimore, Maryland 21218, USA.}
\altaffiltext{2}{Department of Materials Science and Engineering, Johns Hopkins University, Baltimore, Maryland 21218, USA.}
\altaffiltext{3}{Applied Physics Laboratory, Johns Hopkins University, Laurel, Maryland 20723, USA. $^\dagger$Our dear friend and colleague Dr. Bridges passed away on April 26, 2017.}

\begin{article}


\section{Abstract}
To understand the origin of the dunes on Titan, it is important to investigate the material properties of Titan's organic sand particles on Titan. The organic sand may behave distinctively compared to the quartz/basaltic sand on terrestrial planets (Earth, Venus, Mars) due to differences in interparticle forces. We measured the surface energy (through contact angle measurements) and elastic modulus (through Atomic Force Microscopy, AFM) of the Titan aerosol analog (tholin). We find the surface energy of a tholin thin film is about 70.9 mN/m and its elastic modulus is about  3.0 GPa (similar to hard polymers like PMMA and polystyrene). For two 20 $\mu$m diameter particles, the theoretical cohesion force is therefore 3.3 $\mu$N. We directly measured interparticle forces for relevant materials: tholin particles are 0.8$\mathrm{\pm}$0.6 $\mu$N, while the interparticle cohesion between walnut shell particles (a typical model materials for the Titan Wind Tunnel, TWT) is only 0.4$\mathrm{\pm}$0.1 $\mu$N. The interparticle cohesion forces are much larger for tholins and presumably Titan sand particles than materials used in the TWT. This suggests we should increase the interparticle force in both analog experiments (TWT) and threshold models to correctly translate the results to real Titan conditions. The strong cohesion of tholins may also inform us how the small aerosol particles ($\mathrm{\sim}$1 $\mathrm{\mu}$m) in Titan's atmosphere are transformed into large sand particles ($\mathrm{\sim}$200 $\mathrm{\mu}$m). It may also support the cohesive sand formation mechanism suggested by Rubin and Hesp (2009), where only unidirectional wind is needed to form linear dunes on Titan.

\section{Introduction}
Aeolian processes are ubiquitous on bodies with atmospheres (both permanent and ephemeral) in the Solar System, including Earth, Venus, Mars, Saturn's moon Titan (Greeley \& Iversen, 1985), Neptune's moon Triton (Smith et al., 1989) Pluto (Stern et al., 2015) and the comet 67P/Churyumov-Gerasimenko (Thomas et al., 2015). To understand the origin of aeolian processes on Titan, the initiation of saltation has been investigated by measuring fluid threshold wind speed (the lowest wind speed to initiate saltation) using the Titan Wind Tunnel (TWT) (Burr et al., 2015). Complementary to such investigations, the fluid threshold wind speed can be predicted by deriving the force balance of stationary stacking particles. These forces include: the wind drag and lift forces, gravity, and interparticle forces (Shao \& Lu, 2000). During the TWT experiments, the wind drag and lift forces can be manipulated by changing the wind speed and flow regimes in the wind tunnel. The gravity on Titan can be simulated by using lower density material (density\textless$2000\ \mathrm{kg/m^3}$) in the wind tunnel on Earth. However, the interparticle forces are highly dependent on intrinsic material properties (e.g., surface energies). The low density materials used in the TWT (e.g., walnut shells), may have different interparticle forces compared to the real transporting materials on Titan, which are considered to be made of organics deposited from the atmosphere with minor water ice (McCord et al., 2006; Soderblom et al., 2007; Barnes et al., 2008; Clark et al., 2010; Le Gall et al., 2011; Hirtzig et al., 2013; Rodriguez et al., 2014). Thus measurements of the interparticle forces of both the Titan analog materials and the low density materials used in the TWT are necessary, so that we can correctly translate the TWT results to real Titan conditions. 

The formation of dune particles ($\sim$100 $\mathrm{\mu}$m) on Titan is not well understood. Barnes et al. (2015) proposed several mechanisms for haze particles to transform to sand-sized particles: 1) if the sand particles are produced by sintering or by lithification and erosion, then the composition of the sand particles would match the aerosols; 2) if the sand particles are produced by flocculation, the composition of the sand would be similar to the insoluble part of the aerosols in Titan's lakes; 3) the soluble part of the sand particles could form evaporites and the evaporites could be the sand source, too. However, both laboratory and theoretical studies showed that Titan aerosol analogues (`tholin') have low solubility in non-polar solvents (McKay, 1996; Raulin, 1987; Coll et al., 1999; Sarker et al., 2003; Carrasco et al., 2009; He \& Smith, 2014a), which are the major components of Titan's lakes (Brown et al., 2008). Thus, the soluble part of tholin may be a minor composition of Titan's sand. Measuring the interparticle cohesion of these Titan aerosol analog particles could also provide information about the formation of Titan's sand particles.

The interparticle forces consist of van der Waals force, capillary forces due to condensed liquid, and electrostatic forces. In very humid environments, capillary forces usually dominate over the other forces while in low humidity environments, van der Waals forces (solid-solid interaction) dominate at short-range separation. The long-range electrostatic forces may also play an important role in affecting sediment transportation once the particles are placed in motion, thus affecting the impact threshold (the lowest wind speed to maintain saltation, which is usually lower than fluid threshold) more than the fluid threshold.

Apart from intrinsic material properties, interparticle forces are also controlled by environmental conditions, such as relative humidity (RH) and temperature. On Earth, the relative humidity of water generally increases the interparticle forces through capillary condensation (e.g. Jones et al., 2002), while on Titan, the relative humidity of methane or ethane may affect the interparticle forces as well. Temperature also affects the interparticle forces, especially at temperatures near a substance's melting point; a melted quasi-liquid layer could form capillary bridges at surface asperities (the unevenness of surface). For ice in air, this quasi-liquid layer may disappear at around $-20\degree$C (Petrenko \& Whitworth, 1999). Thus near the melting point, the interparticle forces increase with increasing temperature (e.g. Yang et al., 2004; Taylor et al., 2008). 

There are a number of models which describe the adhesion between two smooth surfaces (Maugis, 1992). Two limiting equations are often used to describe the adhesion forces between smooth, dry surfaces. The DMT limit (Derjaguin, Muller, \& Toporov, 1975) generally applies for hard materials and small contacting radii of curvature, and the JKR limit (Johnson, Kendall, \& Roberts, 1971) describes the interparticle adhesion for soft materials and larger contacting radius of curvature. The application of either of the limiting equations depends on the elastic modulus and the surface energy. Here we measured these two material properties for tholin, through contact angle and elastic modulus measurements, thus we can theoretically predict the interparticle forces.

The above theoretical models of interparticle forces usually predict much larger results than found in experimental data. The challenge for these models is the use of over-simplified geometry; actual particles are not usually perfectly round and have asperities to decrease the real contact area. The irregularity and roughness of the particles makes the contact area smaller than if they were perfect molecularly smooth spheres. A number of models have been trying to describe the effect of roughness on adhesion forces (e.g. Greenwood \& Williamson, 1966; Rumpf, 1990; Xie, 1997; Cooper et al., 2000; Rabinovich et al., 2000), however, exact predictions for real particles are still difficult. Thus it is still necessary for us to measure the interparticle forces between actual particles. 

\begin{figure}
\setfigurenum{1}
\begin{center}
\includegraphics[width=20pc]{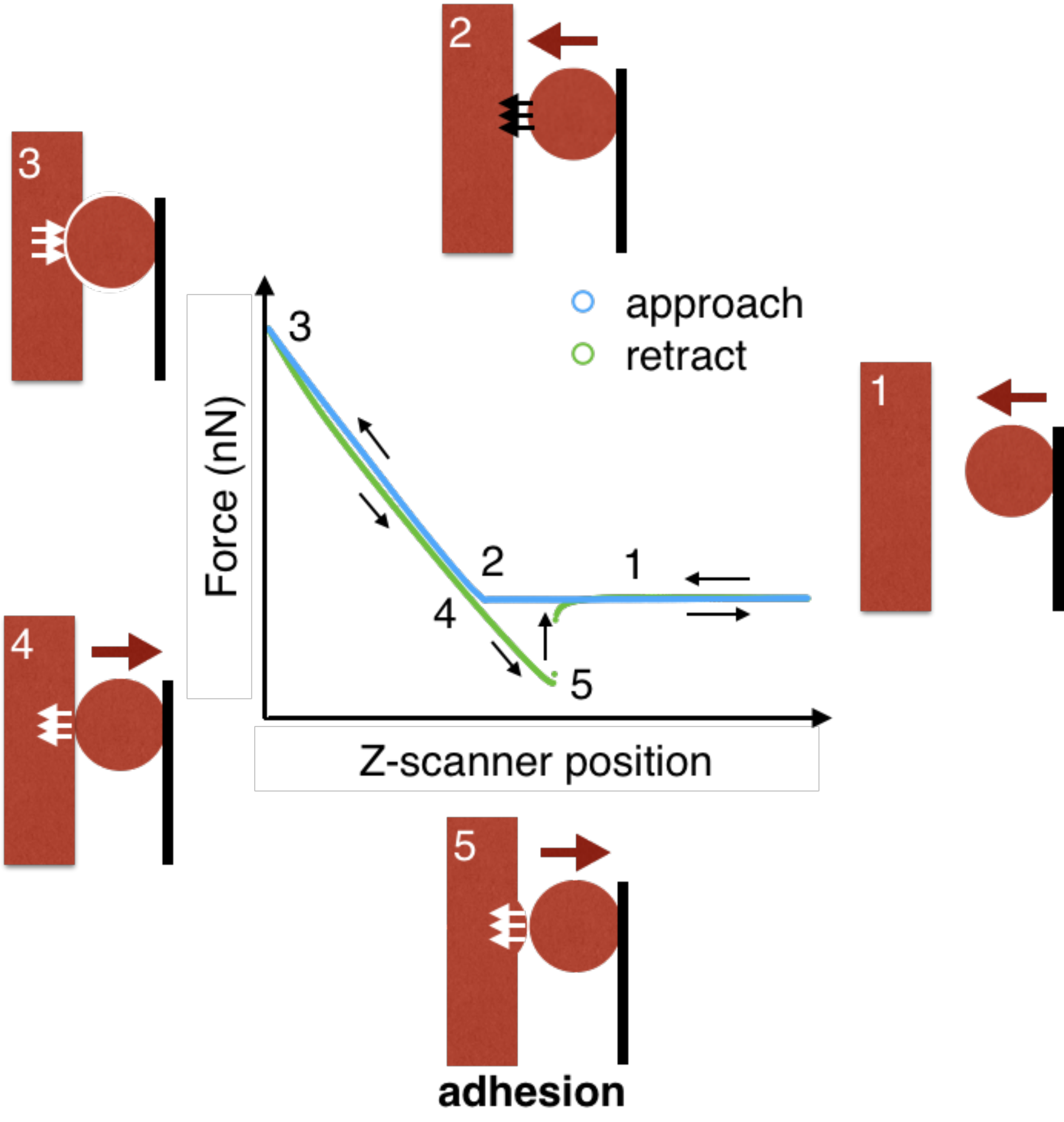}
\caption[]{A typical force curve between a colloid particle and a smooth surface and the particle-surface interactions during different stages of approach and retraction (figure adapted from Haugstad, 2012). (1) The particle is far away enough from surface that there is no interaction force between the particle and surface. (2) The particle contacts the surface which may be proceeded by a slight short range attractive force (in this force curve the attraction is weak). (3) A repulsive force is measured due to the cantilever deflection and possible indentation/compression of the surfaces. (4) The AFM cantilever is retracted but the particle remains in contact. (5) At the final state of contact, the adhesion forces are equal to the maximum pull-off forces of the cantilever. }
\label{fig:pulloff}
\end{center}
\end{figure}

The ability of atomic force microscopy (AFM) to measure forces as a function of surface separation enables us to directly measure the particle-surface or particle-particle separation forces at the single particle level; the forces measured are called the adhesion forces (or pull-off forces) (Ducker et al., 1991). A series of particle-surface interactions during one force-distance curve cycle (approach and retract) are shown in Figure \ref{fig:pulloff}. As the particle approaches the surface, the interaction force increases from zero to attraction between the particle and film. Then the particle may `jump in' to the surface because of the attraction. As the surfaces are pushed together, a repulsive force will be measured. As the AFM cantilever retracts the particle from the surface, the adhesion forces between the particle and the surface will prevent separation. When the pulling force of the cantilever exceeds the maximum adhesion forces, the particle will `jump out' from the surface. Thus the adhesion forces are dependent on the depth of this adhesion minimum. The adhesion forces between single particles reflect a combination of interparticle forces whose relative importance depends on environmental conditions and material properties.

The two limiting equations (DMT and JKR limits) to describe adhesion forces and the derivation of saltation threshold is reviewed below (Section 3). The experimental methods are described in Section 4.1--4.5. In Section 5.1, we used the measured intrinsic material properties: contact angles, surface energy, and elastic modulus of tholin to calculate the theoretical adhesion forces between tholin particles. The measured adhesion forces of AFM silicon tip to several different surfaces are compared in Section 5.2. The results of the adhesion forces measurements for particle-surface and particle-particle interactions are summarized in Section 5.3. To further explore the effect of geometry and environmental conditions on adhesion forces, a tholin coated colloidal particle was used and we measured its adhesion to a flat tholin surface under different relative humidities (Section 5.4). 

\section{Background}

Two simple models are often used to describe the solid-solid interaction forces of smooth, contacting surfaces under dry or low humidity conditions. The DMT model  (Derjaguin, Muller, \& Toporov, 1975) generally applies for hard materials and small contacting radii of curvature, and the interparticle forces can be expressed as:
\begin{equation}
F_{DMT}=2\pi R^*W_A,
\end{equation}
where $\mathrm{R^*}$ is the effective radius of curvature, given by $\mathrm{R^*=(1/R_1+1/R_2)^{-1}}$, where $\mathrm{R_1}$ and $\mathrm{R_2}$ are radii of the contacting particles. $\mathrm{W_A}$ is the work of adhesion; for two solid surfaces made of the same material in vacuum/dry air, 
\begin{equation}
W_A=2\gamma_s
\end{equation}
where $\mathrm{\gamma_s}$ is surface energy of the solid. On the other hand, the JKR model (Johnson, Kendall, \& Roberts, 1971) best describes the interparticle adhesion for soft materials and larger contacting radius of curvature, the interparticle forces for this model are:
\begin{equation}
F_{JKR}=\frac{3}{2}\pi R^*W_A.
\end{equation}
In order to know which model is appropriate for the system we are investigating, we need to calculate an the elasticity parameter $\lambda$ to determine which regime applies (Haugstad, 2012):
\begin{equation}
\lambda=\frac{2.06}{\xi_0}(\frac{R^*\gamma^2_s}{\pi K^2})^{1/3},
\end{equation}
where $\mathrm{\frac{1}{K}=\frac{3}{2}\frac{1-\nu^2}{E}}$, $\nu=0.3$ is the poisson ratio, E is the elastic modulus of the material, and $\mathrm{\xi_0=0.16\ nm}$ is the equilibrium interatomic distance. The DMT model applies when $\mathrm{\lambda<0.1}$ and the JKR model applies when $\mathrm{\lambda>5}$. For tholin, neither its  surface energy ($\gamma_s$) nor its elastic modulus (E) is known, so we cannot predict its interparticle forces under dry conditions. Regardless of which model is applied, the interparticle forces will be a function of particle size and surface energy; thus the uncertainty from the two parameters could strongly affect the calculated theoretical interparticle forces, as well.

The DMT and JKR models only apply for dry or low humidity conditions. At higher RH, when liquid starts to condense on the particles or the surface, the interparticle forces begin to be dominated by capillary forces:
\begin{equation}
F_{capillary}=4\pi R^*\gamma_Lcos\theta,
\label{eq:capillary}
\end{equation}
where $\mathrm{\gamma_L}$ is the surface energy of the condensed liquid and $\mathrm{\theta}$ is the contact angle between the liquid and the solid surface.

Shao and Lu (2000) used the balance of gravity ($F_g\propto d^3$), aerodynamic drag and lift ($\mathrm{F_d}$ and $\mathrm{F_l}$ both $\propto d^2$), and interparticle forces ($F_i\propto d$) to derive the threshold friction wind speed:
\begin{equation}
u^*_{sl}=\sqrt{f(Re^*)(\frac{\rho_p-\rho_a}{\rho_a}gd+\frac{\gamma}{\rho_ad})},
\label{eq:shaoandlu}
\end{equation}
where
\begin{equation}
\gamma=\frac{6}{\pi}\frac{a_i}{a_g}\beta,
\label{eq:gammaandbeta}
\end{equation}
where d is the diameter of the particles, $\mathrm{a_i}$ and $\mathrm{a_g}$ are the moment arm lengths of the interparticle and gravity forces, respectively. The values of f(Re*) and $\gamma$ are acquired by fitting the experimental data from Iversen \& White (1982), where they used a boundary layer wind tunnel to measure threshold wind speed for various materials of different densities (1100--2650 $\mathrm{kg/m^3}$) and sizes (37--673 $\mathrm{\mu}$m). They found f(Re*) is approximately 0.0123, and $\gamma$ is between 1.65--5 $\mathrm{N/m}$ (Shao \& Lu, 2000). The value of $\mathrm{\beta}$ links to the magnitude of the interparticle forces:
\begin{equation*}
F_i=\beta d.
\end{equation*}

Note that the JKR and DMT theories show that $\mathrm{\beta}$ is in the range of $\mathrm{1.5\pi\gamma_s}$ and $\mathrm{2\pi\gamma_s}$. Roughness will further decrease $\beta$. For a 100 $\mathrm{\mu}$m diameter particle, the interparticle forces are estimated to be on the order of $\mathrm{10\ \mu N}$ ($\mathrm{\beta\sim10^{-1}\ N/m}$). However, to fit the experimental threshold wind speed data, Shao and Lu (2000) found out the interparticle forces are only on the order of $\mathrm{10^{-2}\ \mu N}$ ($\mathrm{\beta\sim10^{-4}\ N/m}$), which is not only several orders smaller than the estimated value ($\mathrm{10\ \mu N}$), but also one order smaller than the measured value for quartz sand ($\mathrm{F_i\approx0.1\ \mu N}$, Corn, 1961).

The interparticle forces used in Shao and Lu (2000)'s model are not specific for any particular materials, because the model did not link the range of the parameter $\mathrm{\gamma}$ or $\mathrm{\beta}$ (see Equation {\ref{eq:gammaandbeta}}) to material properties. This might be particularly problematic for Titan because the transporting materials on Titan are mainly organic and their intrinsic interparticle cohesion (the $\mathrm{\beta}$ parameter) could be very different from silicate materials on Earth for which the models were developed.

\section{Methods}
\subsection{Samples and Preparation}
\begin{figure}
\setfigurenum{2}
\begin{center}
\includegraphics[width=25pc]{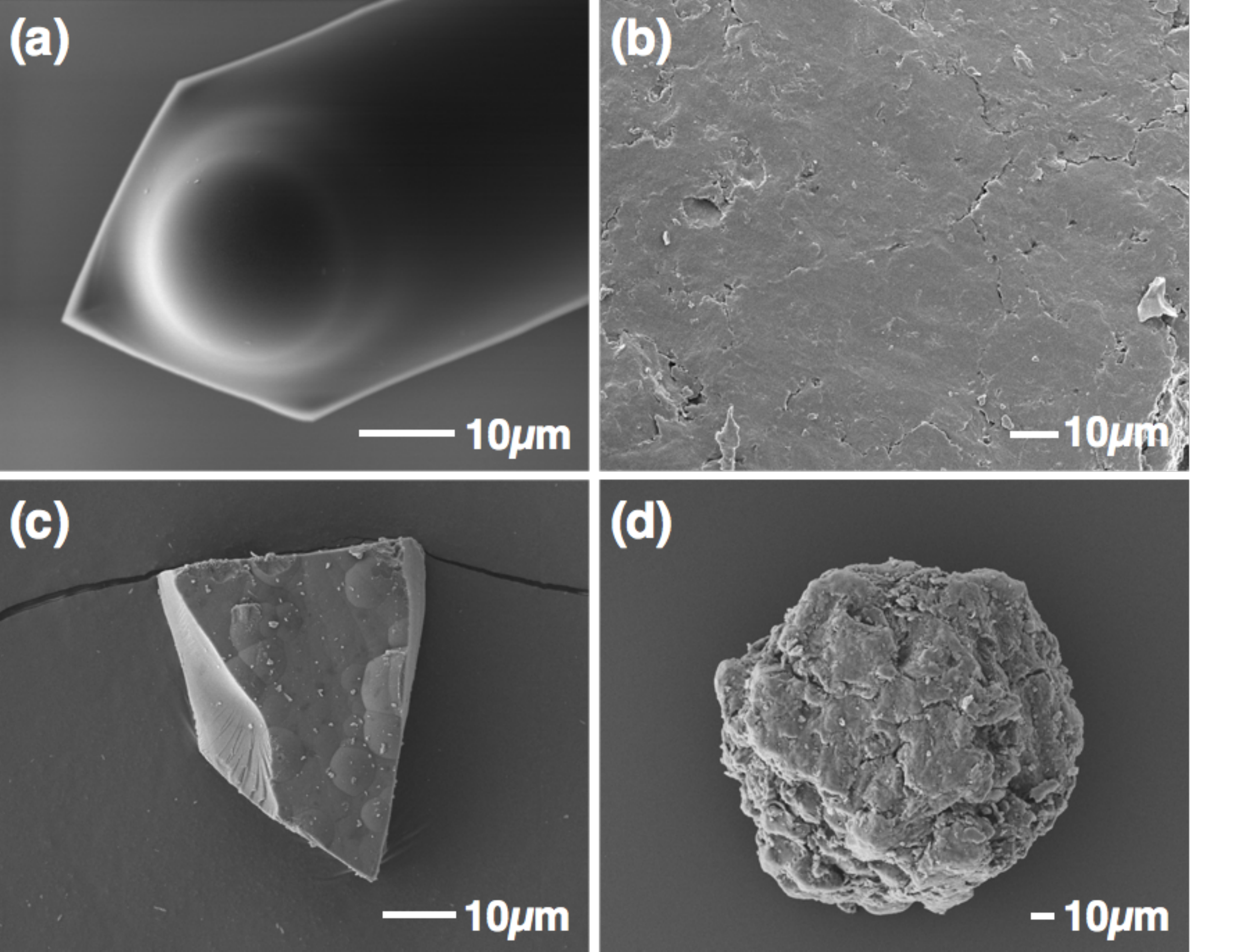}
\caption[]{SEM images of: (a) a tholin-coated colloidal AFM probe (the coated sphere is about 20 $\mu$m in diameter); (b) the flat side of a walnut shell particle (size {$\sim$}800 {$\mu$}m) for tip--surface and particle-surface interactions; (c) a typical tholin particle (size {$\sim$}30 {$\mu$}m) used for particle-particle adhesion forces; (d) a typical walnut shell particle (size 125--150 {$\mu$}m) used for particle-particle adhesion forces.}
\label{fig:tholin_colloid_surface}
\end{center}
\end{figure}

Tholins were produced by exposing 5\% $\mathrm{CH_4}$/$\mathrm{N_2}$ gas mixture to a glow plasma discharge (pressure: 3 Torr, temperature: 100 K), with a 10 sccm flow rate (He et al., 2017). Tholins were deposited: 1) on four mica discs (10 mm diameter), 2) three colloidal probes (AFM cantilevers from sQube with a $\sim$20 $\mu$m diameter borosilicate glass sphere attached to the end of the cantilever), and 3) on the wall of the chamber. The tholin films deposited on mica discs are approximately 1 {$\mu$}m thick, and their RMS roughness is {$\sim$}1 nm. Figure \ref{fig:tholin_colloid_surface}(a) shows a scanning electron microscopy (SEM) micrograph of one of the tholin-coated colloidal probes. Tholin particles deposited on the chamber wall were collected in a dry $\mathrm{N_2}$, oxygen free glove box. The representative Titan Wind Tunnel materials (walnut shells 125--150 $\mu$m), are from the original TWT batches used in Burr et al. (2015) and Yu et al. (2017).

\subsection{AFM and cantilevers}
We used a Bruker Dimension 3100 atomic force microscope. The spring constants of the cantilevers were calibrated by thermal tuning. The spring constant of the regular cantilevers is about 40 N/m and those of the colloidal probes is approximately 2.8 N/m. The sensitivity of the AFM photodiode is measured by indenting a hard surface (cleaved mica sheets, detailed in McGuiggan et al.. 2011).

\subsection{Elastic Modulus Measurements}
We used the AFM as a nanoindenter to measure the stiffness of tholin. We performed two cycles of force-separation curves on a hard surface (silicon, assume no indentation) and on a smooth tholin surface. The z scanner distance (x-axis in Figure \ref{fig:pulloff}) on the hard surface was subtracted from the z scanner distance on the sample surface to get a force (F)--indentation ($\mathrm{\delta}$) curve. The elastic modulus (E) of tholin can then be found as a function of indentation as:
\begin{equation}
E=\frac{3(1-\nu^2)F}{4R^{*1/2}\delta^{3/2}},
\label{eq:elasticmodulus}
\end{equation}
where $\mathrm{\nu=0.3}$ is the Poisson ratio and R* is the radius of the AFM tip ($\sim$10 nm).

\subsection{Contact Angle Measurements}
We performed contact angle measurements on a flat tholin film. We used both a polar (deionized water) and a non-polar (diiodomethane, $\mathrm{CH_2I_2}$) liquids, which usually yields the most reliable surface energy results (Hejda et al., 2010). We also used heptane as an analog to liquid methane or ethane, since the surface tensions are similar (around 20 mN/m). Contact angles were determined by using a Ram\'e-Hart goniometer. 

When a liquid droplet forms on a flat solid surface in an inert atmosphere, we can balance the three phases (liquid, solid, and air) energies by using the Young-Dupr\'e equation:
\begin{equation}
W_{sl}=\gamma_l(1+cos\theta),
\label{eq:youngequation}
\end{equation}
where $\mathrm{\gamma_l}$ is the surface energy of the liquid, $\mathrm{W_{sl}}$ is the work of adhesion (energy to separate the solid and liquid) of the liquid and solid, and $\mathrm{\theta}$ is the contact angle between the liquid-air interface and the solid surface.
Using the geometric mean method, the work of adhesion $\mathrm{W_{sl}}$ can be also approximated as (Owens \& Wendt, 1969):
\begin{equation}
W_{sl}=2(\sqrt{\gamma_s^d\gamma_l^d}+\sqrt{\gamma_s^p\gamma_l^p}),
\label{eq:workofadhesion}
\end{equation}
where $\mathrm{\gamma_s^d}$ and $\mathrm{\gamma_l^d}$ are solid and liquid dispersion contributions to the surface energy, and $\mathrm{\gamma_s^p}$ and $\mathrm{\gamma_l^p}$ are the solid and liquid polar contributions to the surface energy.
When the contact angle measurements are done using two liquids, we have two sets of equations \ref{eq:youngequation} and \ref{eq:workofadhesion} to solve for the surface energy of the solid. A similar harmonic mean method developed by Wu (1971) was also used to calculate the surface energy and the results are similar to the geometric mean method.

\subsection{Adhesion Force Measurements}
\begin{figure}
\setfigurenum{3}
\begin{center}
\includegraphics[width=25pc]{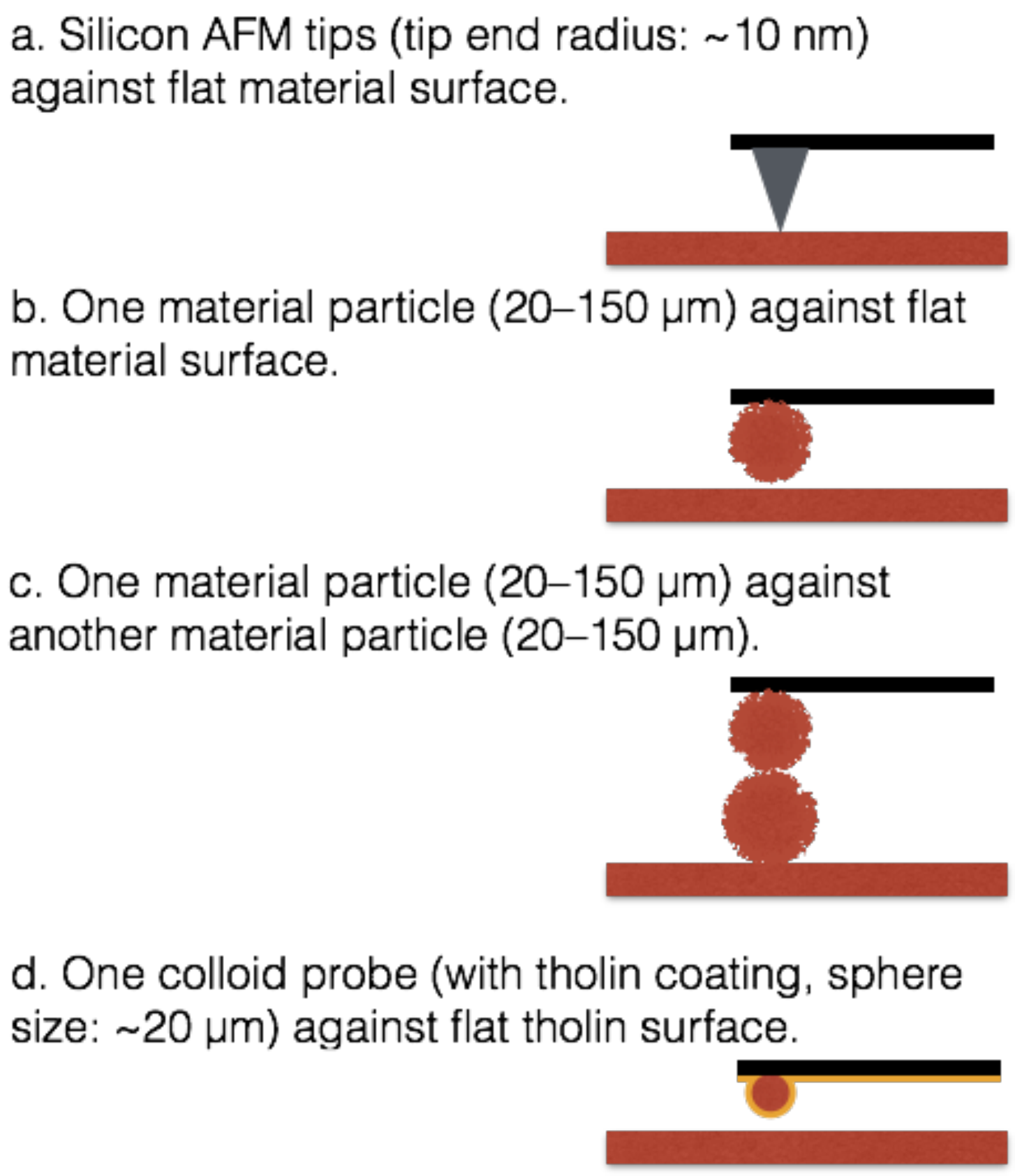}
\caption[]{The four types of AFM setup used for the adhesion measurements. Figure adapted from Jones et al., (2002).}
\label{fig:experiments}
\end{center}
\end{figure}

We performed force-separation curves on four different simple systems to measure adhesion forces, as shown in Figure \ref{fig:experiments}. The adhesion force measurements were all done at a scan rate of 1.5 to 2 Hz ($\approx$4 $\mu$m/s). There was no change in the measured adhesion forces at rates of 0.5 Hz, 5 Hz, and 10 Hz and there was no change in the measured adhesion forces.

To study tip-flat adhesion, we conducted force distance curve with a bare silicon AFM tip to 1) a flat tholin deposited film, 2) a flat quartz surface (Pelco Quartz Substrate from Ted Pella, Inc), and 3) the flat side of a walnut shell particle (size $\sim$800 $\mu$m, see Figure {\ref{fig:tholin_colloid_surface}(b)}), as shown in Figure \ref{fig:experiments}a. 

To study particle-flat and particle-particle cohesion, two kinds of particles were used: a tholin particle ($\sim$30 $\mu$m, see Figure {\ref{fig:tholin_colloid_surface}(c)}) and a walnut shell particle ($\sim$125--150 $\mu$m, see Figure {\ref{fig:tholin_colloid_surface}(d)}). They were glued to AFM cantilevers using epoxy resin. Force curve measurements were conducted for these particles to both flat film (for walnut shell, we used the flat side of an 800 $\mu$m particle) and particles made of the same material as the glued particle, as shown in Figure \ref{fig:experiments}b and \ref{fig:experiments}c. For each particle-flat and particle-particle cohesion measurement, 2--4 spots on the film or 2--4 particles on the substrate were chosen randomly and 6--20 pairs of approach-retract force curves were taken.

To study the variation of adhesion forces with different humidities, we performed the measurements using more controlled contact geometry: a colloidal probe coated with tholin was used as the cantilever. Force curves were obtained between the probe and a flat tholin film, as shown in Figure \ref{fig:experiments}d. We also investigated the effect of relative humidity (RH) on adhesion forces for the tholin coated colloidal probe. The measurements were conducted in a controlled RH environment, varying RH from \textless1\% in a dry nitrogen environment to about 40\% in ambient air. Relative humidity (RH) and temperature were recorded by a digital hygrometer (Dwyer Instrument), the RH range is 0--100\% with an accuracy of $\pm$2\%, and the temperature range is $-$30--85$\degree$C with an accuracy of $\pm$ 0.5$\degree$C.  

\section{Results and Discussion}

\subsection{Intrinsic material properties of tholin and its theoretical adhesion forces}

The contact angle measurements show that for water ($\mathrm{\gamma=72.8\ mN/m}$, $\mathrm{\gamma^d=20.0\ mN/m}$, and $\mathrm{\gamma^p=52.8\ mN/m}$) on a flat thin tholin film, the contact angle is $\mathrm{22\pm5}\degree$. While for diiodomethane ($\mathrm{\gamma=\gamma^d=50.8\ mN/m,\gamma^p=0\ mN/m}$), the contact angle is $\mathrm{50\pm5}\degree$. Thus we can solve Equations \ref{eq:youngequation} and \ref{eq:workofadhesion} for the surface energy of tholin:
\begin{equation}
\gamma_s=\gamma_s^d+\gamma_s^p=(34.3_{-2.9}^{+2.7}+36.6_{-3.8}^{+3.7})\ mN/m=70.9_{-4.8}^{+4.6}\ mN/m
\end{equation}
The contact angle between heptane and tholin is less than $\mathrm{\textless5}\degree$.

From the indentation part of the force curve, we get an elastic modulus (E) of tholin film of about $\mathrm{3.0\pm0.7}$ GPa (Equation \ref{eq:elasticmodulus}), which is consistent with hard polymers like PMMA and polystyrene (Israelachvili, 2011). Using the measured surface energy of tholin $\mathrm{\gamma_s=70.9_{-4.8}^{+4.6}}$ mN/m and its elastic modulus E, we can calculate the elasticity parameter $\lambda$ for an AFM tip or a particle with radius R touching a flat tholin surface:
\begin{equation}
\lambda=\frac{2.06}{\xi_0}(\frac{R\gamma^2_s}{\pi K^2})^{1/3}\approx891R^{1/3}.
\end{equation}
For the tholin particles investigated in this study (both the particle and the coated colloidal particle), $\mathrm{R\approx10\ \mu m}$, thus we get $\mathrm{\lambda\approx19}$, which makes the JKR model most appropriate for this system. The theoretical adhesion force under dry conditions for particle-flat surface system can then be calculated as:
\begin{equation}
F=\frac{3}{2}\pi R^*W_A=6.7\pm0.3\ \mu N.
\label{eq:forcedry}
\end{equation}
For two tholins particle of similar size ($\mathrm{R\approx10\ \mu m}$), the adhesion force is half of the adhesion force for particle-flat system ($\mathrm{3.3\pm0.2}\ \mu$N), since the radius R* is halved for this system.

If particles are exposed in humid terrestrial conditions (water, $\mathrm{\gamma_L=72.8\ mN/m}$), the adhesion forces start to become capillary dominated: 
\begin{equation}
F=4\pi R^*\gamma_{water}cos\theta_{water}=8.5\pm0.3\ \mu N
\label{eq:forcewater}
\end{equation}
If the condensed liquid is liquid methane ($\mathrm{\gamma\approx20\ mN/m}$, Baidakov et al., 2013) as on Titan, then the adhesion forces become:
\begin{equation} 
F=4\pi R^*\gamma_{methane}cos\theta_{methane}\approx2.5\ \mu N,
\label{eq:hydrocarbon}
\end{equation}
here we use the contact angle between tholins and heptane ($\theta\textless5\degree$), leading to $\mathrm{cos\theta=0.996\ to\ 1}$, which is consistent with the value used in Lavvas et al. (2011), $\mathrm{cos\theta}\sim0.995$ between tholin and methane. This value is derived from the experimental results of methane adsorption on tholin films produced in a different laboratory (Curtis et al., 2008).

As RH increases, the particles transition from solid-solid interaction to capillary interactions; there might be a gradual transition for the force (e.g. Christenson, 1988). The capillary condensation that occurs around surface contact sites grow with increasing RH until a liquid film surrounds the macroscopic contact. This transition continues until the capillary meniscus radius exceeds the asperity size, then the interparticle forces are dominated by capillary forces as shown in Equation \ref{eq:capillary} (McFarlane \& Tabor, 1950).

It is interesting to notice that the theoretical interparticle force would actually decrease with increasing liquid methane humidity since the capillary force for liquid methane is lower the van der Waals force under dry conditions. However, if the particles have significant roughness, rather than being perfect spheres, then the van der Waals forces at low humidity would be much smaller. In this case they may be lower than the liquid methane capillary forces, which depend on the capillary radius. 

For comparison, the gravity force for the 20 $\mathrm{\mu}$m size particle is only about  2.9$\times10^{-6}$ $\mu$N to 8.2$\times10^{-6}$ $\mu$N on Titan, using the density range from 500--1400 $\mathrm{kg/m^3}$ (Imanaka et al., 2012; H\"orst \& Tolbert, 2013; He et al., 2017). If the above calculation is done for a tholin particle of radius 100 $\mathrm{\mu}$m (assuming a perfect smooth sphere, which is almost certainly not true for real sand particles on Titan), the contact force is a huge 67 $\mu$N (JKR still applies for tholins of this size), the methane capillary force is 25 $\mu$N, while the gravity force is only 2.9$\times10^{-3}$ $\mu$N to 8.2$\times10^{-3}$ $\mu$N. Since the gravity forces for the sand particles on Titan are so small, the interparticle forces dominate the movement of particles on Titan.

\subsection{Tip to Flat Surfaces Adhesion Forces}
\begin{figure}
\setfigurenum{4}
\begin{center}
\includegraphics[width=25pc]{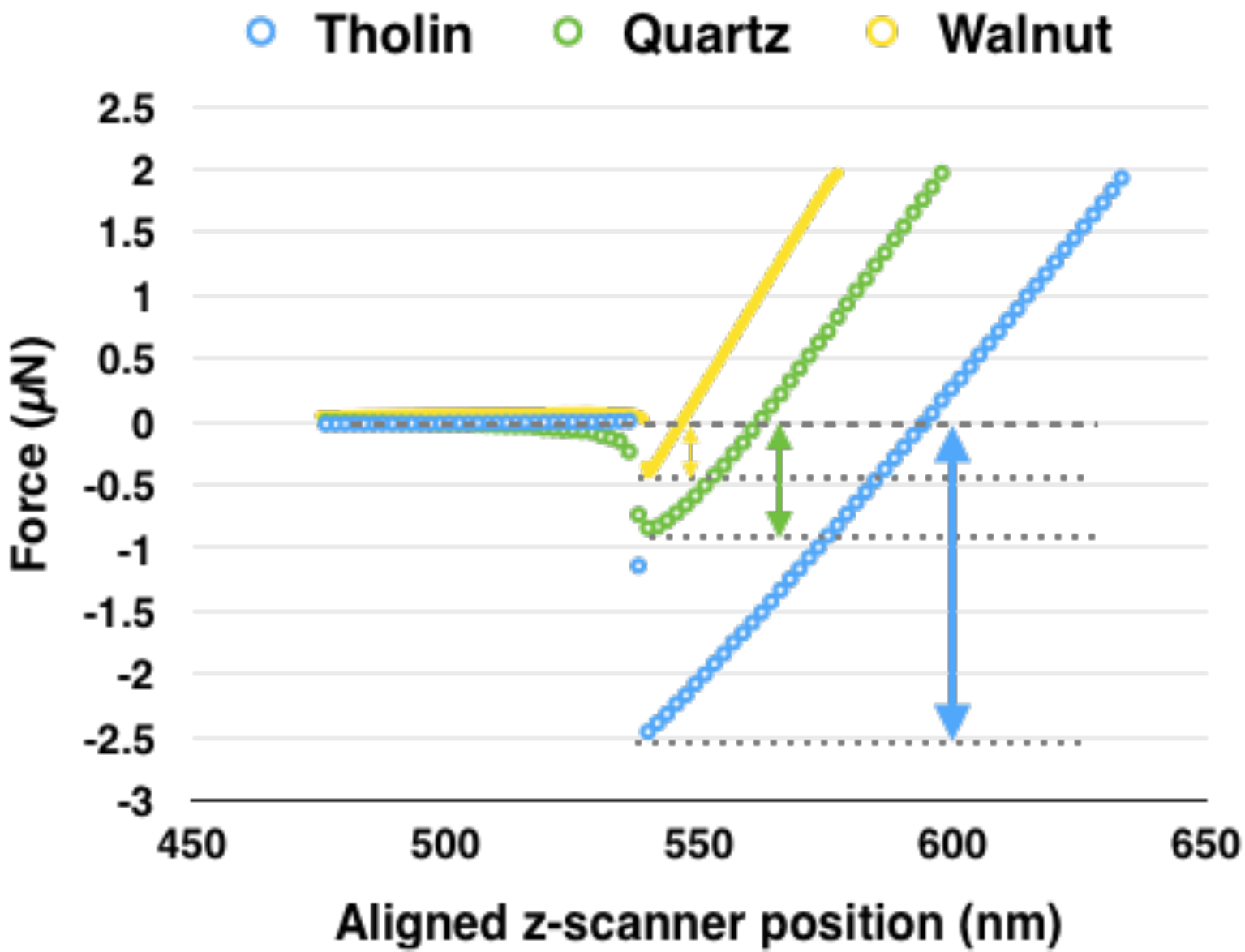}
\caption[]{Retract force curves of a silicon tip to three different flat films, tholin (blue), quartz (green), and the flat side of a 800 $\mathrm{\mu}$m walnut shell (yellow). The maximum pull-off forces are given by the forces at the minimum.}
\label{fig:tip}
\end{center}
\end{figure}

Shown in Figure \ref{fig:tip} are three retraction force curves of a silicon tip (tip radius about 10 nm) to three kinds of flat surfaces: quartz, tholin, and the flat side of a walnut shell particle (800 $\mathrm{\mu}$m). The differences of the adhesion forces are stark between these films. Here the adhesion force is smallest between tip and the flat surface of the walnut shells, about 0.3 $\mu$N, while the adhesion force between silicon tip and tholin film is almost 10 times larger, about 2.4 $\mu$N. The adhesion force between silicon tip and quartz film is in between, about 0.8 $\mu$N.

\subsection{Particle to Particle Adhesion Forces}
\begin{figure}
\setfigurenum{5}
\begin{center}
\includegraphics[width=32pc]{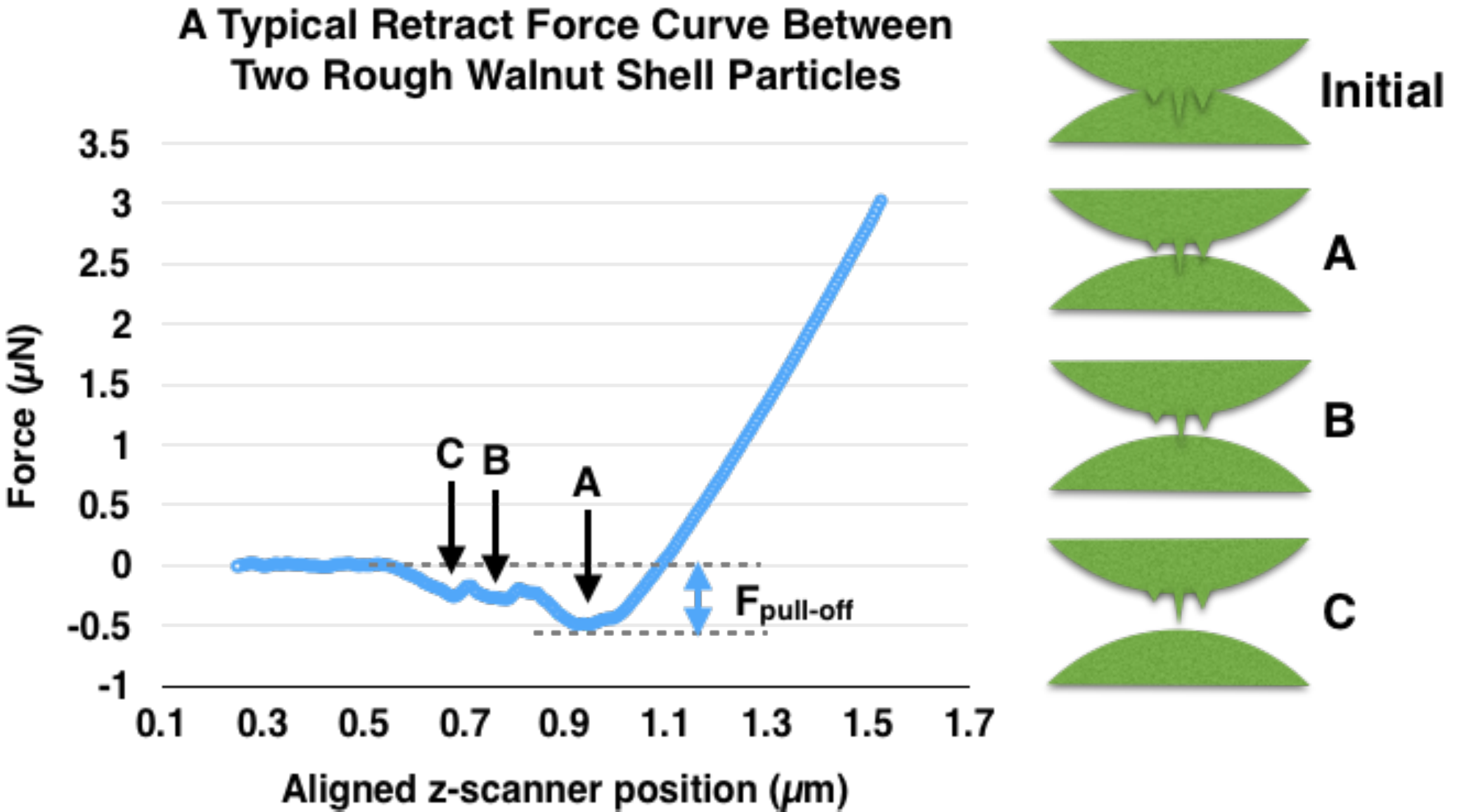}
\caption[]{On the left is a typical retract force curve between two rough walnut shell particles. Multiple pull-off events (three here in this figure marked as A, B, C) were observed. The total pull-off force is the maximum attraction force ({$\mathrm{F_{pull-off}}$}) shown in the figure. For this specific force curve, {$\mathrm{F_{pull-off}=0.5\ \mu N}$}. On the right is an illustration of the three pull-off events due to three different micro- and nano- scale roughness on the surface.}
\label{fig:roughparticles}
\end{center}
\end{figure}

To directly study interparticle cohesion, we attached tholin and walnut particles to the end of the AFM cantilever. The force curves between particles are more complicated since the particles are usually very rough at both micro- and nano- scales, and are not uniformly spherical. We observed multiple pull-off events in our experimental data for some of the very rough particles. An example of a retract force curve between two rough walnut shells particles is shown in Figure {\ref{fig:roughparticles}}. The pull-off force is given by the maximum attraction force as that force is sufficient to pull off all the asperities (Beach et al., 2002). Since we cannot control the surface of the particles used in the wind tunnel, the roughness of the particles will usually lead to high standard deviations in pull-off force measurements because the contact area in every force curve may be different. However, this range of measured values likely represents the actual range of interparticle forces for the wind tunnel experiments.

\begin{figure}
\setfigurenum{6}
\begin{center}
\includegraphics[width=25pc]{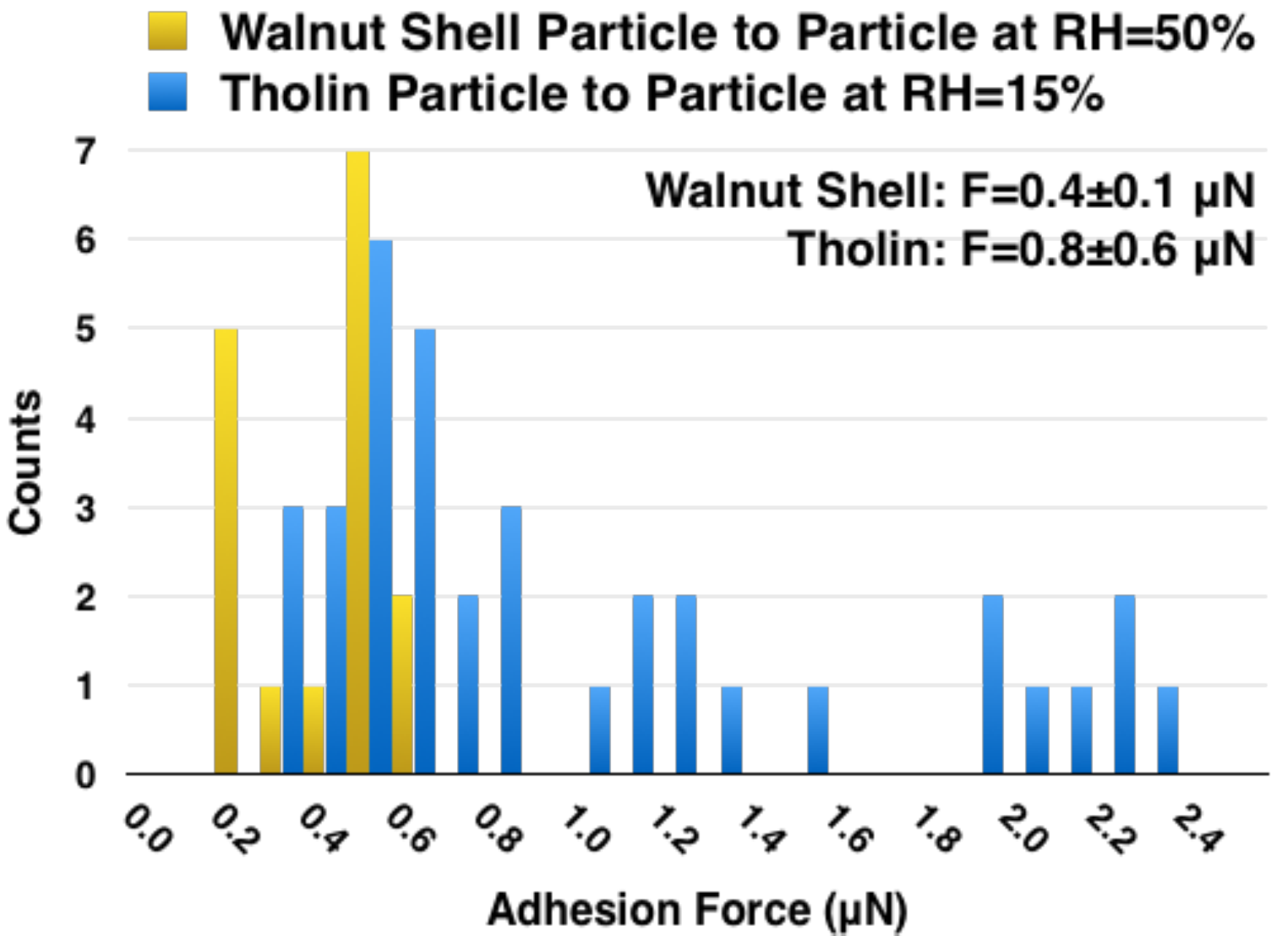}
\caption[]{Histograms of adhesion forces between walnut shell particles (under RH of about 50\%) and tholin particles (under RH of about 15\%). Histograms were used to better characterize the range of the measured interparticle forces.}
\label{fig:particleparticle}
\end{center}
\end{figure}

In Table \ref{table:particletoparticle} and Figure \ref{fig:particleparticle} we show the results of particle-particle adhesion forces for walnut shells and tholins. According to JKR and DMT contact mechanics (Johnson, Kendall, \& Roberts, 1971; Derjaguin, Muller, \& Toporov, 1975), the adhesion forces scale linearly to the particle size. However, measurements show that the smaller 30 $\mathrm{\mu}$m tholin particles actually show larger adhesion forces (average 0.8$\pm$0.6 $\mu$N) compared to the bigger 100 $\mathrm{\mu}$m walnut shells (average 0.4$\pm$0.1 $\mu$N), even under lower humidity (RH$\approx$15\% for tholin measurements and RH$\approx$50\% for walnut shell measurements). This is likely caused by the combined effect of micro and nano surface roughness (walnut shell particle roughness (RMS)$\approx$ 70 nm vs 20 nm for tholin particles) and surface energy ($\mathrm{\gamma_{tholin}\approx70.9\ mN/m}$ and $\mathrm{\gamma_{walnut\ shell}\approx30-50\ mN/m}$, de Meijer et al., 2000). This indicates when calculating the threshold wind speed on Titan, we may need to increase the $\mathrm{\gamma}$ or $\mathrm{\beta}$ values in Equation \ref{eq:gammaandbeta} for Titan sand to accommodate its larger interparticle cohesion. We may also need to incorporate the variability of the measured adhesion forces (caused by surface roughness), shown in Figure \ref{fig:particleparticle} into the threshold model by using a probabilistic distribution of interparticle forces, as suggested by Yang et al. (2004). The large particle-particle adhesion forces of tholin also suggest they are easier to coagulate to form larger particles, and this may provide insight on how the small aerosol particles in Titan's atmosphere are transformed to large sand-sized particles on Titan's surface, if that is indeed the sand formation mechanism. The higher cohesion of tholin particles may also support the alternative formation mechanism of Titan's linear dunes, where Rubin and Hesp (2009) suggests that only unidirectional wind is needed for strong-cohesive sand to form longitudinal dunes.

\begin{table}
\centering
\caption{Adhesion forces between different particles}.
\vspace{0.2cm}
\begin{tabular}{|c|c|c|c|c|c|}
\hline
Particle & Size & Surface & Adhesion & Std. Dev. & RH range\\
Name & ($\mathrm{\mu}$m) &  & ($\mu$N) & ($\mu$N) & (\%)\\
\hline
\multirow{2}{*}{Walnut Shell} & \multirow{2}{*}{125} & Walnut Shell Particle A & 0.3 & 0.1 & \multirow{2}{*}{50.1--57.0}\\
&  & Walnut Shell Particle B & 0.4 & 0.1 & \\
\hline
Walnut Shell & 125 & Walnut Shell Film & 0.5 & 0.3 & 50.1--57.0\\
\hline
\multirow{3}{*}{Tholin} & \multirow{3}{*}{30}& Tholin Particle A & 0.6 & 0.4 & \multirow{3}{*}{14.7--15.9}\\
&  & Tholin Particle B & 0.5 & 0.1 &\\
&  & Tholin Particle C & 1.5 & 0.6 &\\
\hline
Tholin & 30 & Tholin Coated Surface & 6.7 & 5.1 & 14.7--15.9\\
\hline
Tholin Coated & \multirow{2}{*}{20}& \multirow{2}{*}{Tholin Coated Surface} & 2.9 & 1.2 & 44.3--50.7\\
Colloid &  &   & 1.2 & 0.6 & 23.8--24.3 \\
\hline
 &  &   & 2.6 & 0.1 & 1.7--1.9\\
 &  &   & 2.9 & 0.1 & 5.9--6.3 \\
Tholin Coated& \multirow{2}{*}{20} & {Tholin Coated Surface}  & 3.1 & 0.1 & 10.3--10.5 \\
Colloid &  & {$\mathrm{N_2}$ Enclosed--Location 1}  & 3.1 & 0.1 & 20.0--20.2 \\
 &  &   & 3.3 & 0.1 & 29.8--30.0 \\
 &  &   & 5.0 & 0.2 & 35.0--35.3 \\
 &  &   & 5.3 & 0.3 & 39.6--40.0 \\
\hline
 &  &   & 2.2 & 0.1 & 0.8--1.0\\
 &  &   & 3.9 & {\textless}0.1 & 5.7--5.9 \\
Tholin Coated& \multirow{2}{*}{20} & {Tholin Coated Surface}  & 3.2 & 0.1 & 11.0--11.2 \\
Colloid &  & {$\mathrm{N_2}$ Enclosed--Location 2}  & 5.1 & 0.2 & 20.2--20.5 \\
 &  &   & 5.1 & 0.1 & 30.3--30.6 \\
 &  &   & 6.5 & 0.2 & 40.7--40.9 \\
\hline
\end{tabular}
\vspace{-0.2cm}
\label{table:particletoparticle}
\end{table}

\subsection{Adhesion Forces of a Tholin Coated Sphere to Flat Tholin Surfaces}
To further investigate the effect of humidity on adhesion forces, we changed the AFM experimental setup to the configuration shown in Figure \ref{fig:experiments}d. Instead of measuring the interaction between two irregularly shaped tholin particles, we used a tholin-coated spherical particle and a flat tholin film. This allows us to reduce the effect of surface roughness and focus on the effect of RH on adhesion forces. 

\begin{figure}
\setfigurenum{7}
\begin{center}
\includegraphics[width=30pc]{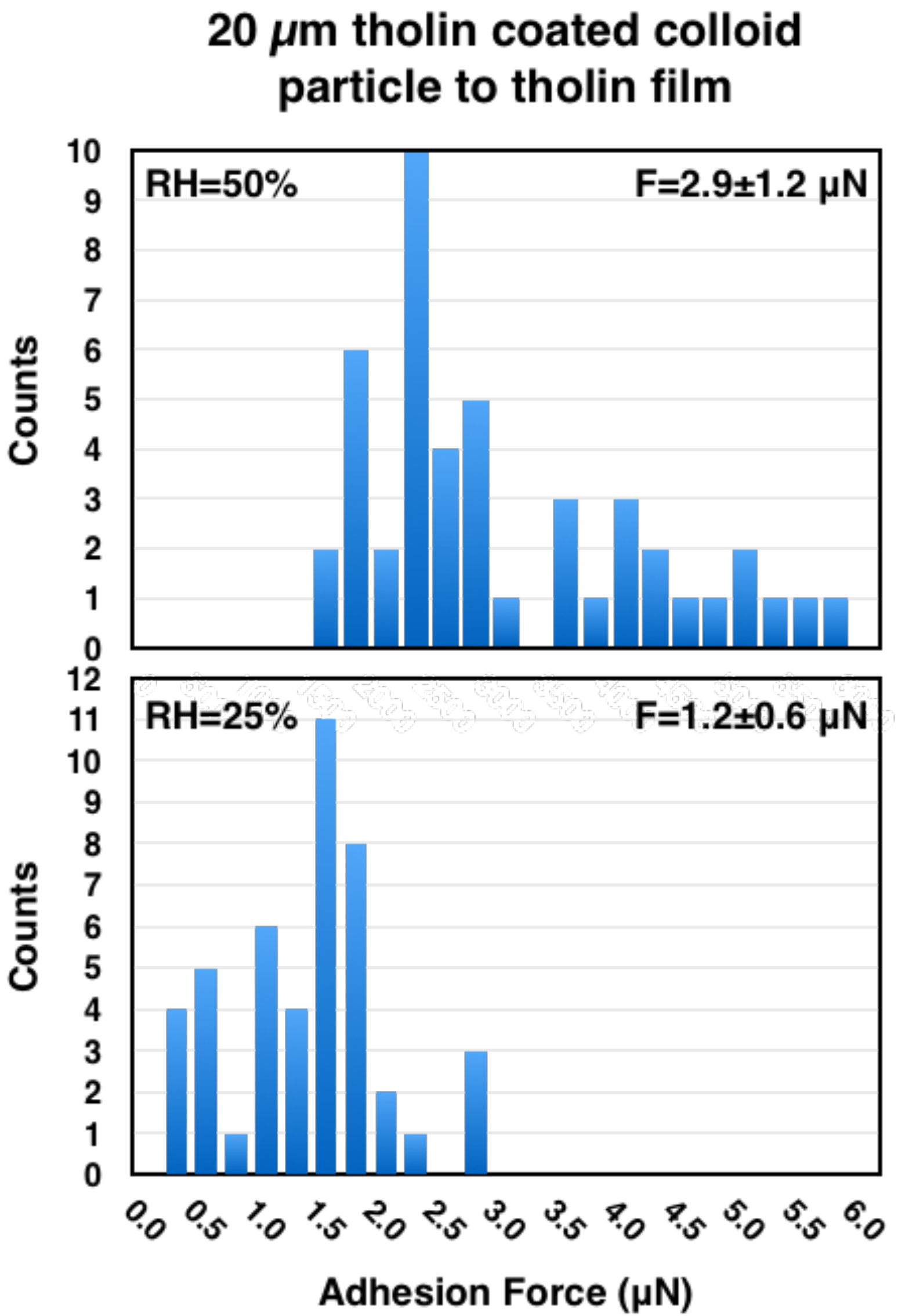}
\caption[]{Histograms of adhesion forces for tholin coated 20 $\mu$m colloid probe to tholin film at RH=50\% and 25\%.}
\label{fig:tholincolloid}
\end{center}
\end{figure}

We first obtain forces curves with the tholin-coated particle on different positions of a tholin film at two different RH values, RH=25\% and RH=50\%. The distribution of pull-off forces is shown in Figure \ref{fig:tholincolloid} and the values are also summarized in Table \ref{table:particletoparticle}. The standard deviation of the measurements is generally large, likely due to the differences in surface roughness between the positions. However, even with the measured standard deviation, the adhesion force differences between the two RHs are significant: at RH=50\%, the measured adhesion forces are almost double the adhesion forces at RH=25\%. This suggests that the adhesion forces between tholin particles are strongly affected by water vapor in air. 

\begin{figure}
\setfigurenum{8}
\begin{center}
\includegraphics[width=35pc]{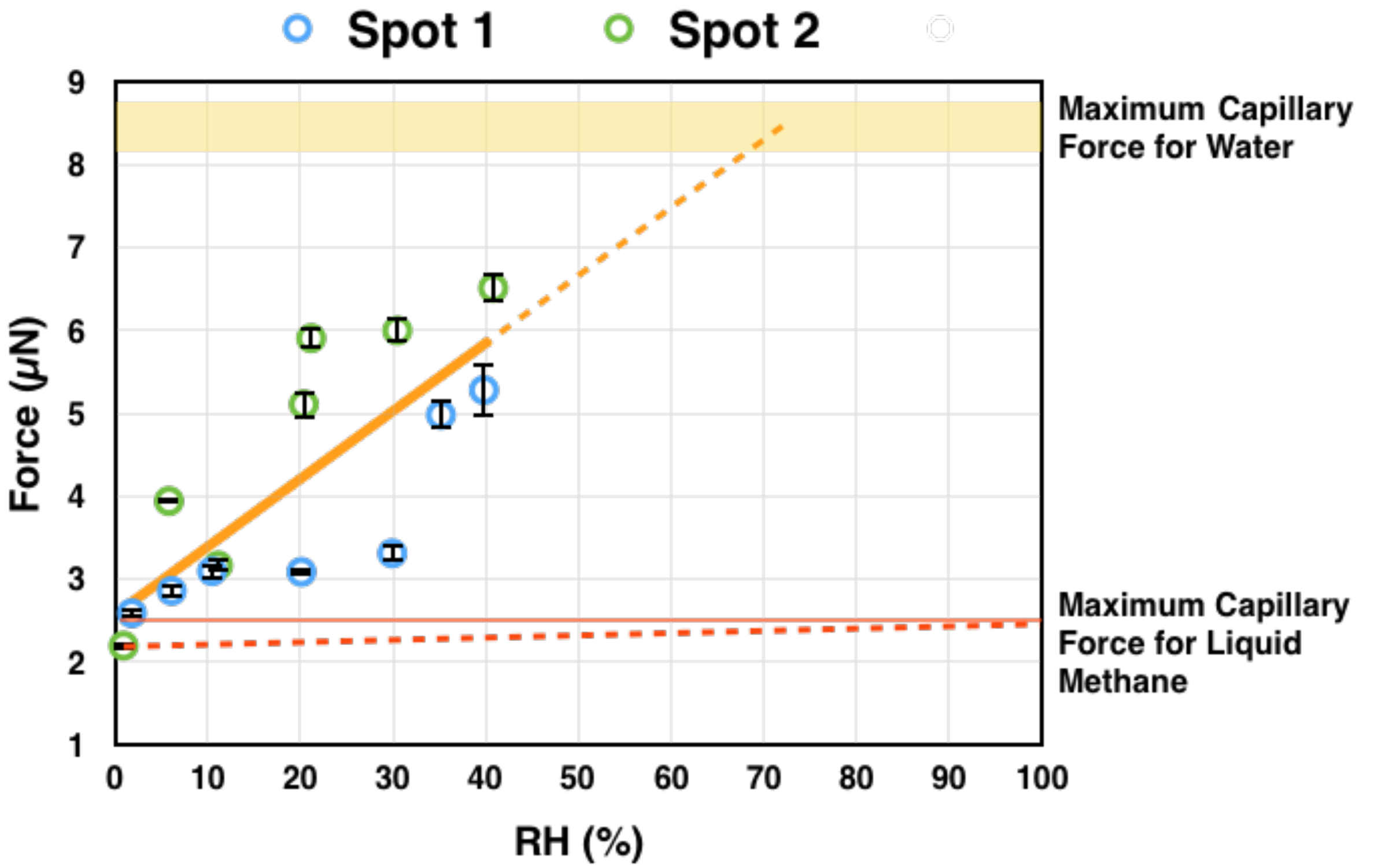}
\caption[]{Shown here are the adhesion forces vs relative humidity (RH) of water between tholin coated sphere and a flat tholin surface. Blue circles and green circles were obtained at two different locations on the tholin film. The solid orange line shows a linear fit to the data for both locations. The orange stripe corresponds to the theoretical capillary force (including error bars) for water vapor (see Equation \ref{eq:forcewater}), and the red thin stripe corresponds to the theoretical capillary force for liquid methane vapor (including error bars also, see Equation \ref{eq:hydrocarbon}). The orange dashed line and the red dashed line show the extrapolated force-RH relationship for water and for liquid methane, respectively.}
\label{fig:force-rh}
\end{center}
\end{figure}

Since there is very little water vapor in Titan's atmosphere, we want to know the adhesion forces for tholin under dry conditions. To do that, we enclosed the entire AFM system in a dry nitrogen glove bag and again measured the interaction force between a tholin coated colloidal probe and a tholin surface. We also measured the adhesion forces as the RH was increased, which are shown in Figure \ref{fig:force-rh}. The adhesion forces are still strong under very low RH (RH\textless1\%), at around 2--2.5 $\mu$ N. The forces are lower than the theoretical adhesion forces under dry conditions, 6.7 $\mu$N (see Equation \ref{eq:forcedry}), which is likely due to nanoscale surface roughness on the coated tholin surfaces ($\mathrm{roughness\ (RMS)\approx}$1 nm for tholin film). As RH increases, the adhesion gradually increases until ambient humidity (RH=40--50\%). This is consistent with the behavior of a hydrophilic surface (Jones et al., 2002), which can be explained by the existence of abundant polar molecules in tholins (He et al., 2012; He \& Smith, 2014b). The theoretical maximum capillary force (Equation \ref{eq:forcewater}) for water vapor is shown as the dashed line in Figure \ref{fig:force-rh}; the extended linear fit to the data would reach this force at around RH=70--80\%. If we assume the forces would change from pure van der Waals contact force to capillary forces for different liquid methane humidities, similar to water, we can draw an imaginary Force--RH trend line for the forces to reach to full capillary forces (liquid methane, Equation \ref{eq:hydrocarbon}) with liquid methane humidity increasing. As discussed in Section 5.1, because of the surface roughness of real particles, the van der Waals force under dry conditions is lower than theoretical calculations. Thus our results suggest that with increasing liquid methane humidity, the interparticle forces will actually increase. One major limitation on our understanding is that contact forces are strongly affected by surface roughness, and we do not currently know roughness of the sand particles on Titan. Future missions to the surface are required to assess this important parameter.

Temperature may also play a role in adhesion forces. First, as temperature decreases, the surface tension of the liquid decreases, thus decreasing the capillary forces. Second, if the substance is close to its melting point, a quasi-liquid layer could cause additional capillary forces. Then as the temperature decreases away from the melting point, the adhesion forces would drop because the capillary forces decrease or disappear. The measured surface tension of liquid methane is $\sim$20 mN/m at Titan's surface temperature (Baidakov et al., 2013). We used this value to make theoretical predictions of capillary forces for liquid methane (Equation \ref{eq:hydrocarbon}). Tholin is a stable solid at room temperature ($\sim$300 K), and tholins do not appear to melt to temperatures of at least $\sim$350 K (He \& Smith, 2014a). Thus room temperature measurements should not be affected by the potential additional capillary forces formed by the quasi-liquid layers resulting from melting, and the adhesion forces under 94 K should be similar to the forces at room temperature.

\section{Conclusion}
This study shows the first direct measurements of adhesion forces between tholin particles. It indicates that the threshold wind speed on Titan could be larger than reported in Burr et al., (2015), since the interparticle forces between tholin particles are much larger than walnut shells used in the TWT. Measurements of the adhesion between a tholin coated colloidal probe and a flat tholin surface show a significant adhesion force even under Titan conditions (with no water vapor), which also indicates that the small Titan aerosol particles ($\sim$1 $\mathrm{\mu}$m) could coagulate efficiently into larger sand-sized particles. The high cohesiveness of tholin may also support the alternative formation mechanism of Titan's dunes, where only unidirectional wind is required with cohesive sediment to form longitudinal dunes on Titan.

\section{Acknowledgements}
Data of the measured adhesion forces are available in this paper and in supplements. Partial support to X. Yu and N. Bridges is provided by NASA grant NNX14AR23G/118460 that was selected under the Outer Planets Research Program (Bridges PI). C. He is supported by the Morton K. and Jane Blaustein Foundation. P. McGuiggan is supported by the 3M Nontenured Faculty Grant and the National Science Foundation (NSF) under Grant CMMI-0709187. We also gratefully appreciate the contribution from the late Dr. Nathan Bridges.

\end{article}
\clearpage

\end{document}